\documentclass{aastex631}
\usepackage{multirow}
\usepackage{booktabs}
\usepackage{gensymb}
\usepackage{enumerate}
\usepackage{url}
\usepackage{ulem} 

\usepackage{booktabs}
\usepackage{threeparttable}

\newcommand{\Teff}{$\emph{T}\mathrm{_{eff}}$}

\shorttitle{Super Li-rich Unevolved Stars}
\shortauthors{T.-S. Yan et al.}

\begin{document}

\title{Discovery of Nine Super Li-rich Unevolved Stars from the LAMOST Survey}
\author[0000-0003-2318-9013]{T.-S. Yan}
\affiliation{CAS Key Laboratory of Optical Astronomy, National Astronomical Observatories, Chinese Academy of Sciences, Beijing 100101, China}
\affiliation{School of Astronomy and Space Science, University of Chinese Academy of Sciences,  Beijing 100049, China}
\author[0000-0002-0349-7839]{J.-R. Shi}
\email{sjr@nao.cas.cn}
\affiliation{CAS Key Laboratory of Optical Astronomy, National Astronomical Observatories, Chinese Academy of Sciences, Beijing 100101, China}
\affiliation{School of Astronomy and Space Science, University of Chinese Academy of Sciences,  Beijing 100049, China}
\author{L. Wang}
\affiliation{National Astronomical Observatories/Nanjing Institute of Astronomical Optics \& Technology, Chinese Academy of Sciences, Nanjing 210042, China}
\affiliation{CAS Key Laboratory of Astronomical Optics \& Technology, Nanjing Institute of Astronomical Optics \& Technology, Chinese Academy of Sciences, Nanjing 210042,  China}
\author[0000-0002-8609-3599]{H.-L. Yan}
\affiliation{CAS Key Laboratory of Optical Astronomy, National Astronomical Observatories, Chinese Academy of Sciences, Beijing 100101, China}
\affiliation{School of Astronomy and Space Science, University of Chinese Academy of Sciences,  Beijing 100049, China}
\author[0000-0002-1619-1660]{Z.-M. Zhou}
\affiliation{Department of Astronomy, Beijing Normal University Beijing 100875, China}
\author[0000-0002-4391-2822]{Y.-T. Zhou}
\affiliation{Department of Astronomy, School of Physics, Peking University, Beijing 100871, China}
\affiliation{Kavli Institute for Astronomy and Astrophysics, Peking University, Beijing 100871, China}
\author[0000-0003-3240-1688]{X.-S. Fang}
\affiliation{CAS Key Laboratory of Optical Astronomy, National Astronomical Observatories, Chinese Academy of Sciences, Beijing 100101, China}
\author[0000-0002-6647-3957]{C.-Q. Li}
\affiliation{CAS Key Laboratory of Optical Astronomy, National Astronomical Observatories, Chinese Academy of Sciences, Beijing 100101, China}
\affiliation{School of Astronomy and Space Science, University of Chinese Academy of Sciences,  Beijing 100049, China}
\author[0000-0002-6448-8995]{T.-Y. Chen}
\affiliation{CAS Key Laboratory of Optical Astronomy, National Astronomical Observatories, Chinese Academy of Sciences, Beijing 100101, China}
\affiliation{School of Astronomy and Space Science, University of Chinese Academy of Sciences,  Beijing 100049,  China}
\author[0000-0002-4440-4803]{X.-J. Xie}
\affiliation{CAS Key Laboratory of Optical Astronomy, National Astronomical Observatories, Chinese Academy of Sciences, Beijing 100101, China}
\affiliation{Science School, Tibet University, Lhasa, Tibet 850000, China}

\begin{abstract}
Based on the Large Sky Area Multi-Object Fiber Spectroscopic Telescope (LAMOST) medium-resolution spectroscopic survey (MRS), we report the discovery of nine super Li-rich unevolved stars with A(Li) $>$ 3.8\,dex. These objects show unusually high levels of lithium abundances up to three times higher than the meteoritic value of 3.3\,dex, which indicates that they must have experienced a history of lithium enrichment. It is found that seven of our program stars are fast rotators with $vsini>9$ km\,s$^{-1}$, which suggests that the accretion of circumstellar matter may be the main contributor to the lithium enhancement of these unevolved stars, however, other sources cannot be excluded.
\end{abstract}

\section{Introduction} \label{sec:introduction}
Lithium (Li) is one of the key chemical elements that link cosmology to the structure and evolution of stars, and its complex evolution history remains a mystery \citep{DelgadoMena2015,Guiglion2016,Grisoni2019,Randich2020,Matteucci2021}. The information of its evolutionary is of great benefit to the study of many key questions in astrophysics, such as the early Universe, the formation and evolution of the Milky Way, the chemical enrichment of stellar populations, and the structure and internal processes of stars \citep{Randich2021}.

Li is the only element that can be produced at three extremely different astrophysical sites \citep{Charbonnel2021}. In addition to the Big Bang nucleosynthesis, the other two sites are the spallation processes of the Galactic cosmic-ray particles on interstellar matter nuclei \citep{Prantzos2012,Olive1992}, and the nucleosynthesis at specific phases of the stellar evolution. Stars that have evolved to a specific stage to contribute Li including core-collapse supernovae \citep[SNe,][]{Woosley1990,Kusakabe2019}, novae \citep{Hernanz1996,Jose2002,Tajitsu2015}, red giant branch (RGB) stars \citep{Sackmann1999}, asymptotic giant branch (AGB) stars  \citep{Nollett2003,Sackmann1992} and active stars \citep{Kelly2020}.
The primordial Li abundance predicted by the baryon-to-photon ratio from the Planck satellite \citep{Pitrou2018} is A(Li)\footnote{A(Li) = log[N(Li)/N(H) + 12] where N is the number density of atoms. } $\sim$ 2.7\,dex, which is based on the Standard Big Bang Nucleosynthesis model.

It is found that there are some unevolved stars with abnormally high Li abundance, a few of which even largely exceeds the meteoritic value of A(Li) = 3.3\,dex. 
\citet{Deliyannis2002} discovered, for the first time, a dwarf (J37) with A(Li) = 4.29\,dex in NGC 6633 \citep[also see,][]{Laws2003, Ashwell2005}, and \citet{Koch2011} found the turnoff star 
J1741-5343 with $\mathrm{A(Li)}$ of 4.21\,dex in the old metal-poor globular cluster NGC 6397 \citep[also see,][]{Koch2012}. Recently, \citet{Li2018} reported the discovery of two very 
metal-poor subgiants ([Fe/H]$<-$1.8\,dex) with A(Li) = 4.55\,dex (J0741+2132) and 3.45\,dex (J0758+4703), respectively.

The stars with unusually high Li abundance suggest that they should experience Li enrichment. However, the current models on the Li enhancement mechanism remain controversial, and none of them can well explain those stars. In the case of J37,  \citet{Deliyannis2002} suggested that the action of diffusion is responsible for the enhancement of Li, while \citet{Ashwell2005} and \citet{Laws2003} proposed that the accretion of circumstellar matter is the best explanation for its abundance anomaly. For J1741-5343, the high Li abundance can be explained by the interaction with an RGB companion \citep{Koch2011, Koch2012}.

At the moment, only a few super Li-rich unevolved stars have been found, therefore, more such stars are needed to investigate the mechanism of Li enhancement, especially from large spectroscopic survey as pointed out by \citet{Koch2012}. The Large Sky Area Multi-Object Fiber Spectroscopic Telescope survey \citep[LAMOST,][]{Cui2012,Deng2012,Zhao2012,Yan2022} can play an important role in this area due to its unparalleled ability in gathering spectra. In this letter, we report nine unevolved stars from the LAMOST medium-resolution spectroscopic survey (MRS). In Sect.~\ref{sec:data}, we describe the selection of super Li-rich candidates. The determination of the Li abundances and the rotation velocities is presented in Sect.~\ref{sec:abun}. The mechanisms of Li enrichment are discussed in Sect.~\ref{sec:diss}, while the conclusions are given in Sect.~\ref{sec:conclusions}.

\section{Selection of the super Li-rich candidates}\label{sec:data}

The LAMOST spectroscopic survey has provided us with massive medium-resolution spectra, which are collected by the innovative active reflecting Schmidt telescope located in Xinglong Observatory, China \citep{Cui2012}. The optical system has an effective aperture of 3.6\,m to 4.9\,m with a wide field of view of $\sim$5$^{\circ}$. The spectroscopic system contains 16 spectrographs with 32 integrated 4K$\times$4K CCD cameras, and a total of 4,000 spectra can be obtained simultaneously via 4,000 fibers. The LAMOST MRS survey includes blue and red bands, and their wavelength ranges are 4950$-$5350\,\AA\ and 6300$-$6800\,\AA\, respectively, with a resolution power of R $\sim$ 7500.

Combining the information of the stellar atmospheric parameters provided by the LAMOST data release 7 (DR7) \citep[LRS,][]{Luo2015}, and the Li abundances derived from the LAMOST medium-resolution spectra by \citet{Gao2021}, we pick out nine unevolved stars with very strong Li lines in their spectra as the super Li-rich candidates from 125,436 objects with surface gravity log~$\emph{g}\geqslant$ 3.5\,dex.

\section{Stellar Parameters and Li Abundances}\label{sec:abun}

\subsection{Stellar atmospheric parameters}
The stellar atmospheric parameters (effective temperature {\Teff}, log~$\emph{g}$ and metallicity [Fe/H]) used in this analysis are taken from the LAMOST-LRS \citep{Luo2015}, while the microturbulent velocities ($\xi$) are adopted based on a relation to the effective temperature \citep[see][]{Gao2021}. All the information is presented in Table~\ref{tab:parameter}.

\begin{deluxetable*}{cccccccccc}
\tablenum{1}
\tablecaption{The stellar atmospheric parameter of our program stars\label{tab:parameter}}
\tabletypesize{\small}
\tablehead{
\colhead{UCAC4} & \colhead{{\Teff}} & 
\colhead{ log\,$\emph{g}$} & \colhead{[Fe/H]} & 
\colhead{$\xi$} & \colhead{{\Teff, RM05}} & 
\colhead{{\Teff, C10}} & \colhead{log\,$\emph{g}$$\mathrm{_{plx}}$} & 
\colhead{(V$-$K$_{\rm S}$)} & \colhead{E(V$-$K$_{\rm S}$)} \\ 
\colhead{} & \colhead{(K)} & \colhead{(dex)} & \colhead{(dex)} & \colhead{(km\,s$^{-1}$)} & 
\colhead{(K)} & \colhead{(K)} & \colhead{(dex)} &
 \colhead{(mag)} & \colhead{(mag)}
} 
\startdata
440-009448 & 5105 & 3.94 &  0.09 & 1.1 & 4664 & 4739 & 4.43 & 2.781 & 0.29 \\
441-011058 & 5508 & 4.07 &  0.17 & 1.1 & 5188 & 5262 & 3.94 & 2.321 & 0.37 \\
451-011087 & 5541 & 4.23 &  0.16 & 1.1 & 5192 & 5266 & 4.22 & 2.060 & 0.11 \\
569-010383 & 5800 & 4.14 &  0.27 & 1.2 & 5520 & 5585 & 4.33 & 2.532 & 0.86 \\
596-052739 & 5611 & 4.00 & -0.20 & 1.1 & 5707 & 5801 & 4.05 & 1.535 & 0.00 \\
606-009417 & 5552 & 4.10 &  0.14 & 1.1 & 4598 & 4667 & 4.46 & 4.413 & 1.83 \\
629-030411 & 6807 & 3.79 &  0.22 & 1.9 & 6992 & 7030 & 4.09 & 1.578 & 0.74 \\
646-050572 & 5781 & 4.10 & -0.69 & 1.2 & 5697 & 5816 & 3.95 & 1.541 & 0.00 \\
677-057614 & 5754 & 4.01 & -0.09 & 1.2 & 5656 & 5745 & 3.98 & 1.570 & 0.00 \\
\enddata
\end{deluxetable*}

It is known that Li abundance is sensitive to {\Teff} \citep{Spite1982}, therefore, it is necessary to evaluate the reliability of {\Teff}. To verify this, we also estimate {\Teff} from the color index (V$-$K$_{\rm S}$)$_{0}$ using the temperature scales of \citet{Ramirez2005} and \citet{Casagrande2010}. The magnitudes of V and K$_{\rm S}$ are taken from UCAC4 \citep{Zacharias2013} and 2MASS \citep{Cutri2003}, respectively, while the reddening corrections are obtained  from the online 3D dust map\footnote{\url{http://argonaut.skymaps.info}} provided by \citet{Green2019} using the geometric distances of \citet{Bailer-Jones2021}. 
The effective temperatures derived from the color index are shown in Table~\ref{tab:parameter}.

Comparing to the effective temperatures derived from LAMOST-LRS, a consistent result can be found for the four low reddening stars, while the other five stars, i.e. UCAC4\, 440-009448, 441-011058, 451-011087, 569-010383 and 606-009417, show lower {\Teff} estimated from the color index, especially for UCAC4\, 606-009417, the difference is as large as $\sim$1,000\,K. This star is identified as young stellar object (YSO), and locates in the Perseus OB2 association \citep{Azimlu2015}. In addition, UCAC4\, 440-009448, 441-011058 and 451-011087 have also been identified as YSOs, and all of them are in the Orion Star-forming Complex region \citep{Nakano1999,Kounkel2018,Zari2018}. As to UCAC4 569-010383, according to its position and distance, it locates in the Taurus-Auriga complex region \citep{Krolikowski2021}. As noted by \citet{Xiang2021}, such stars should suffer larger reddening than those of the 3D ``average'' dust map of \citet{Green2019}. For examples, \citet{Kunder2017} noted that the effective temperature of UCAC4\, 441-011058 derived from LAMOST-LRS is $\sim$300 K higher than that estimated by the infrared flux method. UCAC\, 606-009417 has an E(V$-$K$_{\rm S}$) of 1.83 from \citet{Green2019}, while \citet{Schlafly2014} and \citet{Schlegel1998} suggested that its reddening is E(V$-$K$_{\rm S}$) $\sim$ 2.62 and $>$ 2.91, respectively. Thus, we rather adopt the $T_{\rm eff}$ from LAMOST-LRS.

\begin{figure}[ht]
\figurenum{1}
\centering
\includegraphics[scale=2.2]{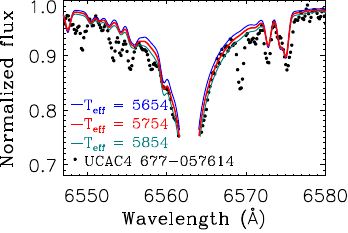}
\caption{The observed spectrum (black dots) around H$\alpha$ line, superimposed on the synthetic spectra. 
\label{fig:halpha}}
\end{figure}

To further demonstrate that the adopted {\Teff} is reliable, we fit the H$\alpha$ absorption lines for five objects (the remaining four stars are YSOs, two of which show H$\alpha$ emission lines and the other two show weak H$\alpha$ absorption lines, Table~\ref{tab:abundance}). As an example, in Fig.~\ref{fig:halpha},  we show the fitting result for UCAC4\, 677-057614, and it can be seen that the observed H$\alpha$ line wing can be well fitted. 

We also calculate the log\,$\emph{g}$ based on the parallax from \emph{Gaia} EDR3 \citep{GaiaCollaboration2021}, and present them in Table~\ref{tab:parameter} (log\,$\emph{g}_\mathrm{plx}$). The discrepancy between the log\,$\emph{g}$ obtained from LAMOST-LRS and  those from the parallax is less than 0.5\,dex, which has no obvious impact on the derived Li abundances.

\subsection{Li abundnaces}

The Li abundances of all our program stars are derived with the spectrum synthesis method, we adopt the MAFAGS-OS model atmospheres \citep{Grupp2004,Grupp2009} and the atomic line data near the Li resonance line at 6708\,\AA\  from \citet{Carlberg2012} to calculate the theoretical profiles with an interactive IDL code Spectrum Investigation Utility \citep[SIU,][]{Reetz1991}. For the NLTE calculation, we adopt the Li atomic model from \citet{Shi2007}. 

\begin{figure*}[ht]
\figurenum{2}
\centering
\includegraphics[scale=1.9]{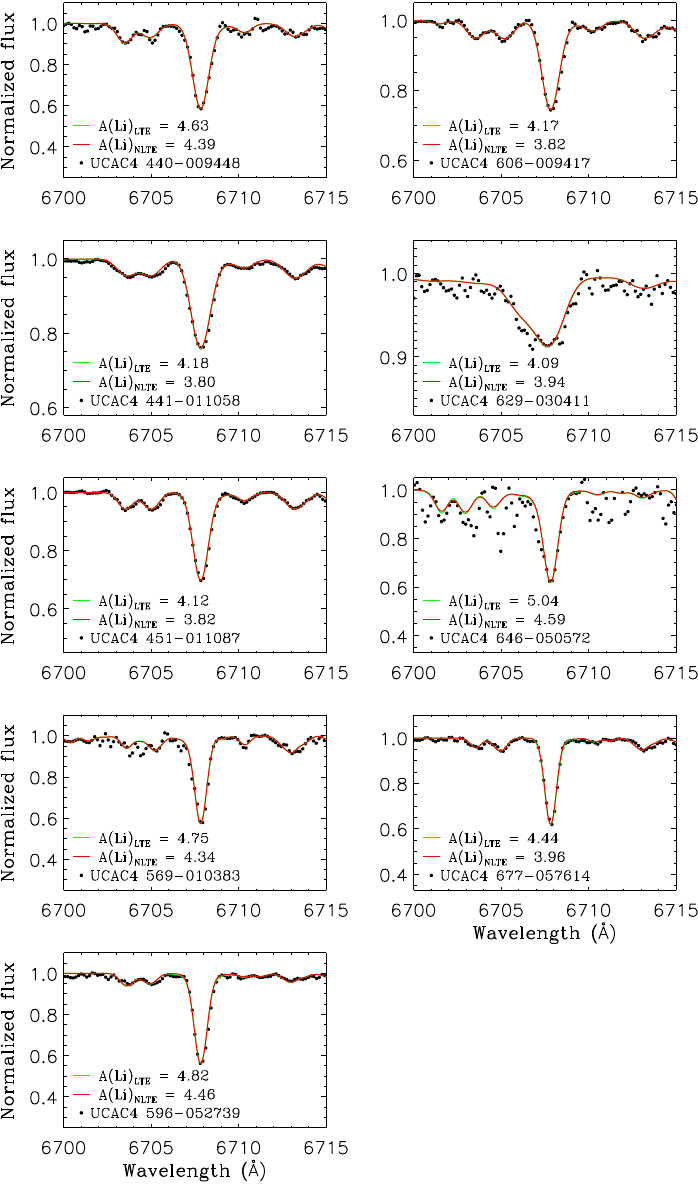}
\caption{The observed spectra (black dots) around the Li line at 6707.8 {\AA}, superimposed on the synthetic spectra under LTE (solid green lines) and NLTE assumptions (solid red lines).\label{fig:fitting}}
\end{figure*}

It is noted that the non-local thermodynamic equilibrium (NLTE) effects of the Li resonance line can not be ignored for Li-rich stars, and it can be larger than 0.2\,dex \citep{Li2018,Yan2018}. Recently, \citet{Wang2021} noted that the 3D-NLTE corrections of the Li resonance line can be up to $\sim$0.15\,dex more negative with respect to previous estimates. Both LTE and NLTE Li abundances are derived with the resonance line at 6708\,\AA, and the results are shown in Table~\ref{tab:abundance}. Fig.~\ref{fig:fitting} presents the fitting result for our program stars. All the nine program stars have NLTE Li abundances higher than 3.8\,dex, which are more than three times of the meteoritic value (3.3\,dex). We note that the NLTE effects are very large for our program stars, with the largest one reaching $\sim$0.5\,dex. 

The uncertainties of Li abundances are mostly due to the errors of temperature, thus we estimate this effect by increasing {\Teff} by 200\,K \citep[two times of the typical uncertainties,][]{Luo2015}, and show them in Table~\ref{tab:abundance}. The equivalent widths (EWs) of Li resonance lines are also shown in Table~\ref{tab:abundance}. For each object, we derive EW by the theoretical line profile synthesized from its Li abundance.

\begin{deluxetable*}{rrrrrrrrrr}
\tablenum{2}
\tablecaption{Li abundances and rotation velocities of our program stars. \label{tab:abundance}}
\tabletypesize{\small}
\tablehead{
\colhead{UCAC4} & \colhead{$\mathrm{A(Li)_{LTE}}$} & 
\colhead{$\mathrm{A(Li)_{NLTE}}$} & \colhead{EW} & 
\colhead{$v\sin{i}$} & \colhead{$M/M_\odot$} & 
\colhead{$R/R_\odot$} & \colhead{$P_\mathrm{rot}$} & 
\colhead{$v_\mathrm{rot}$} & \colhead{class} \\ 
& \colhead{(dex)} & \colhead{(dex)} &\colhead{(m{\AA})} &   
\colhead{(km s$^{-1}$)} &&  & \colhead{(days)} & \colhead{(km\,s$^{-1}$)} 
} 
\startdata
440-009448 & 4.63 $\pm$ 0.27 & 4.39 $\pm$ 0.24 & 496.6  & 18  & 0.78 & 0.89 & 1.69  & 26.6 & YSO \\
441-011058 & 4.18 $\pm$ 0.24 & 3.80 $\pm$ 0.21 & 317.7  & 26   & 1.38 &  2.51 & 3.55 & 35.8 & YSO \\
451-011087 & 4.12 $\pm$ 0.24 & 3.82 $\pm$ 0.20 & 324.2  & 20   & 1.04 & 1.50 & 3.70 & 20.5 & YSO \\
569-010383 & 4.75 $\pm$ 0.23 & 4.34 $\pm$ 0.22 & 372.3  & 9     &  1.16 & 1.56 & ...     & ...     & ...      \\
596-052739 & 4.82 $\pm$ 0.15 & 4.46 $\pm$ 0.20 & 427.5  & 8     & 0.96 & 1.67 & ...     & ...     & ...     \\
606-009417 & 4.17 $\pm$ 0.24 & 3.82 $\pm$ 0.21 & 330.0 & 25   & 1.00 & 0.90 & ...     & ...     & YSO \\
629-030411 & 4.09 $\pm$ 0.16 & 3.94 $\pm$ 0.15 & 163.7 & 45   & 1.60 & 1.89 & ...     & ...     & ...     \\
646-050572 & 5.04 $\pm$ 0.24 & 4.59 $\pm$ 0.20 & 435.0 & 11   & 0.82 & 1.58 & ...     & ...     & ...     \\
677-057614 & 4.44 $\pm$ 0.23 & 3.96 $\pm$ 0.27 & 327.2 &  $<$6   & 1.03 & 1.71 & ...     & ...     & ...     \\
\enddata
\end{deluxetable*}

\subsection{Projected rotation velocities}
Projected rotation velocity ($v\sin{i}$) is an important indicator for diagnosing the mechanism of Li enrichment. We take full width of the half maximum (FWHM) of the arc lamp line near 6708\,\AA\ as the width of instrument profile, and the macro-turbulence velocity is derived from the empirical relation to {\Teff} from \citet{Gray1984}. The derived $v\sin{i}$ are presented in Table \ref{tab:abundance}.
\subsection{TESS Light Curves}\label{sec:light}
\begin{figure*}[htbp]
\figurenum{3}
\centering
\includegraphics[width=15cm]{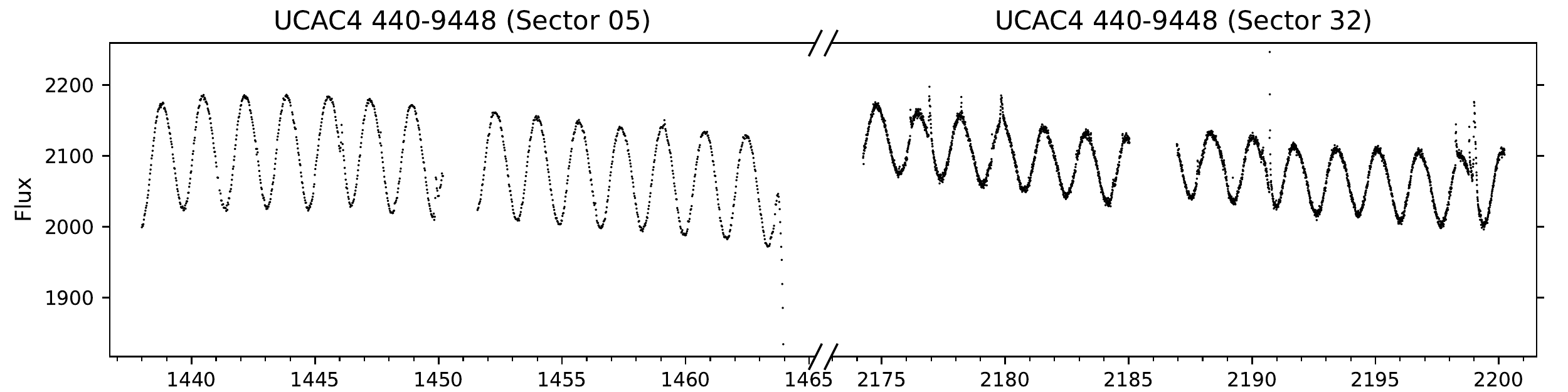}
\includegraphics[width=15cm]{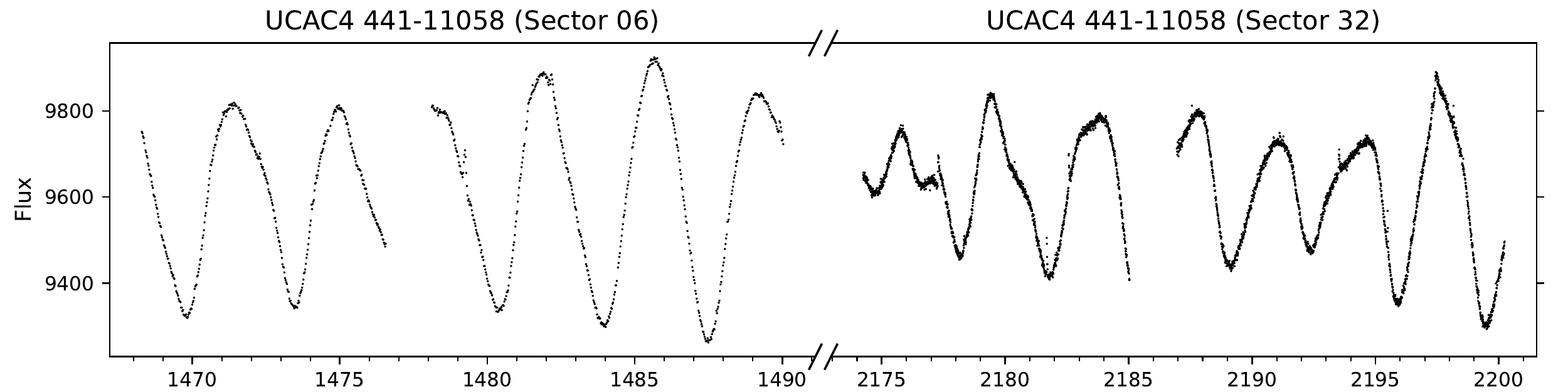}
\includegraphics[width=15cm]{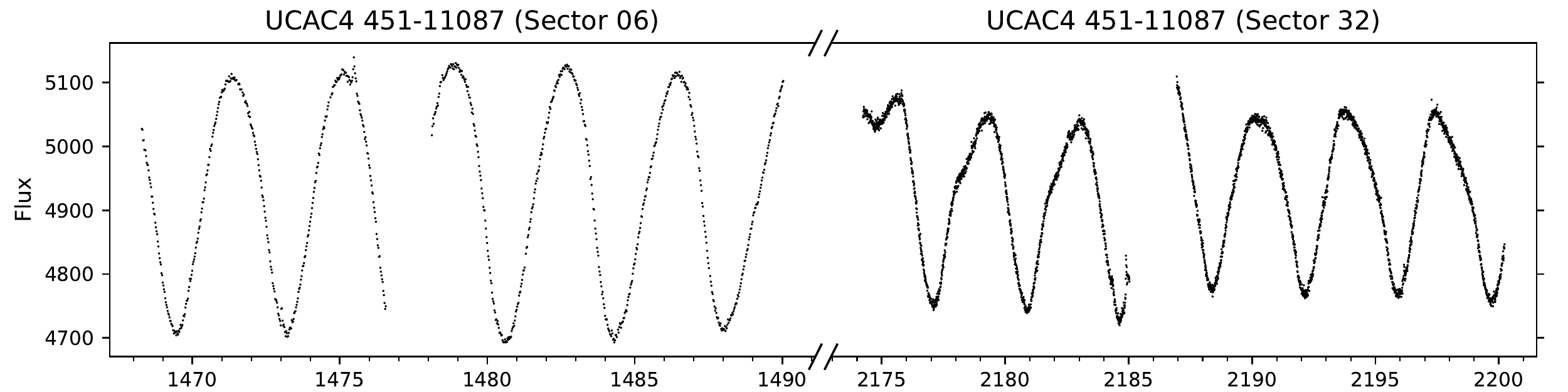}
\includegraphics[width=15cm]{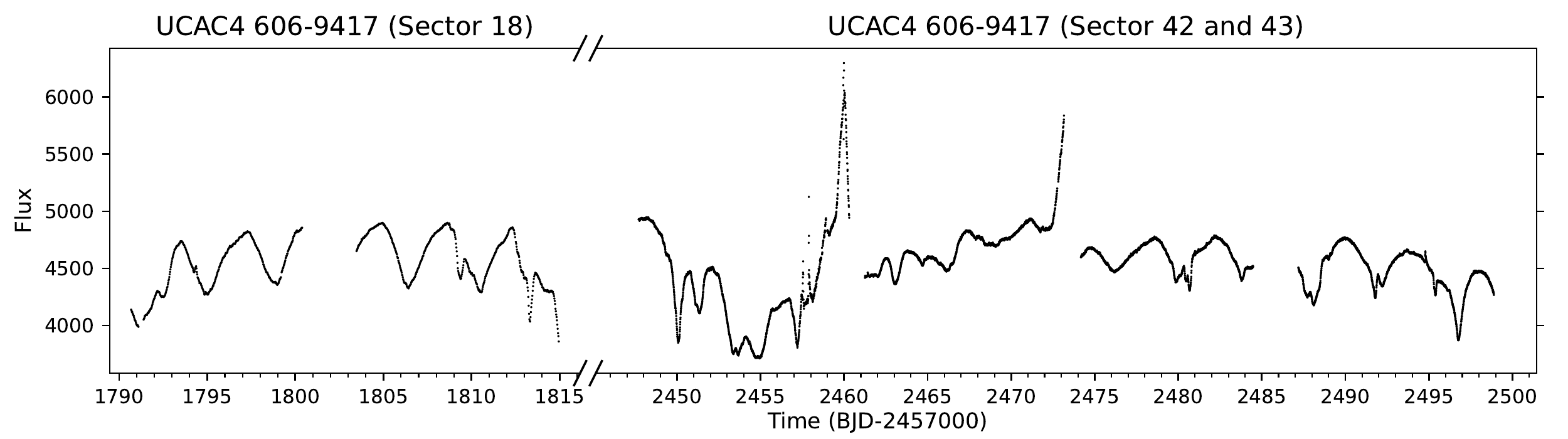}
\caption{Light curves of the four sample stars that exhibit periodic or semi-regular brightness changes. Light curves are extracted using simple aperture photometry from the \emph{TESS} full frame images. Each sector covers $\sim$27 days.}\label{fig:tesslc}
\end{figure*}

We extract the light curves of nine program stars from the Transiting Exoplanet Survey Satellite (\emph{TESS}, \citet{Ricker2014}) full-frame images (FFIs). We cross-match our program stars with the TESS Input Catalog (TIC), and download the time series FFIs of 15 $\times$ 15\,pixels around each target using the TESSCut service\footnote{\url{https://mast.stsci.edu/tesscut/}} on MAST archive. We perform simple aperture photometry with either 3$\times$3 or 5$\times$5\,pixels (depending on the $T$ magnitude) centered on each target, and subtract the background light determined from the pixels outside a certain radius for any known sources in the field. We find that all the nine stars are covered by at least one sector in \emph{TESS}, and four of them show obvious periodic or semi-regular brightness changes. The sector-by-sector light curves are plotted in Fig.~\ref{fig:tesslc}, and the periods range from 1.6 to 4.0 days with amplitudes between 5 --10\%. The flux variations are most likely caused by the rotational modulation of stellar spots for three of them. Therefore, we derive their rotational periods ($P_\mathrm{rot}$) by finding the peaks on the Generalized Lomb-Scargle periodogram \citep{Zechmeister2009} of the light curves. The results are shown in Table~\ref{tab:abundance}. We further derive the rotational speed ($v_\mathrm{rot}$) using the relation of $v_\mathrm{rot} = 50.58(R_*/R_\odot)/P_\mathrm{rot}$, where $R$ is the stellar radius, $P_\mathrm{rot}$ is the rotational period in the unit of days, and $v_\mathrm{rot}$ is in the unit of km\,s$^{-1}$. Considering the upper limit of projected rotational velocities $v\sin{i}$ derived from LAMOST-MRS, we find all the values of $v\sin{i}$ are smaller or agree well with the $v_\mathrm{rot}$ from light curves, which meets our expectation. The stellar radii ($R_*$) and masses ($M_*$) are derived by interpolating the YaPSI evolution tracks \citep{Spada2017}.

The light curve of UCAC4\, 606-009417 has a complicated pattern, which is different from those of other three stars. This star shows clear periodic flux variations in most of time during \emph{TESS} Sector 18, whereas exhibits some irregular dips in brightness with a time-scale of $\sim$1\,day and depths of 3--12\%. We search the \emph{Gaia} EDR3 dataset to check any common proper-motion (CPM) companions closed to this star, and find that it has a CPM companion with a projected separation of 4''.1 and $\Delta G$=4.55. The similar proper motions indicate they are moving along nearly the same direction in the sky, and their parallaxes ($\varpi=3.3925 \pm 0.0185$\,mas and 3.2477 $\pm$ 0.0982\,mas) are consistent within 1.5$\sigma$. Therefore, they are probably gravitationally bounded. The projected separation corresponds to $1240\pm37$ AU, suggesting the system is a wide binary with G7V+M1V components.
\section{Discussion}\label{sec:diss}
\begin{figure}[ht]
\figurenum{4}
\centering
\includegraphics[scale=1]{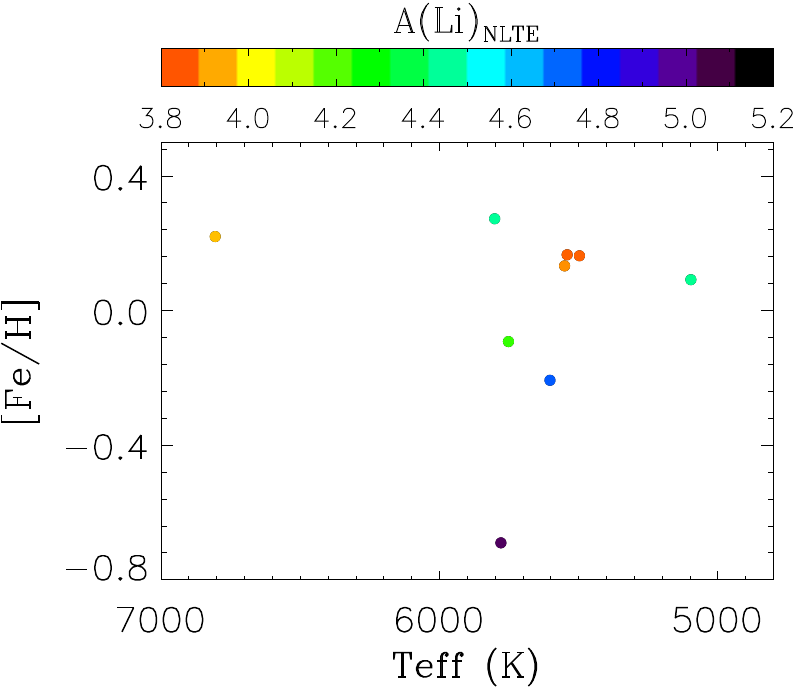}
\caption{Effective temperature versus metallicity for the nine super Li-rich unevolved stars. Stars are color-coded according to their NLTE Li abundances.} \label{fig:parameter}
\end{figure}

We present  the distribution of  the nine super Li-rich unevolved stars in the metallicity$-${\Teff} plane color-coded according to their NLTE Li abundances in Fig.~\ref{fig:parameter}.  Eight of them locate within a small range of the solar metallicity ($-$0.4\,dex $<$ [Fe/H] $< +$0.4\,dex), while the most metal-poor one is UCAC4\, 646-05072 with [Fe/H] of $-0.69$\, dex. Except for the hottest star UCAC4\, 629-030411 with {\Teff} of 6,807\,K, the  effective temperatures of others are between 5,100\,K to 5,800\,K.

We check the variation of radial velocities (RVs) of each object from its consecutively observed spectra and the light curves from TESS of all our program stars, and find no obvious evidence of binaries except UCAC4\,606-009417. 

Although the mechanisms of Li enrichment for the super Li-rich unevolved stars, have been discussed widely \citep{Deliyannis2002, Laws2003, Ashwell2005, Koch2011, Koch2012}, they are still unclear. In addition to the complex origin of Li \citep{Randich2020}, another major factor is that such peculiar objects are so rare. Therefore, it is important to discuss the origin of these super Li-rich unevolved stars: 

Four program stars, i.e., UCAC4\, 440-009448, 441-011058, 451-011087 and 606-009417, are YSOs \citep{Azimlu2015,Nakano1999, Kounkel2018, Zari2018}. The H$\alpha$ lines of four objects show emission or weak  absorption.  It is suggested by \citet{Romano2021} that in the inner Galaxy of $R_{\rm GC}\sim$4\,kpc, the Li abundances can reach as high as 4.0\,dex at the current time, however, the observed upper envelope Li abundance for the young star formation regions are around 3.3\,dex. The NLTE Li abundances of the four young stars are higher than 3.8\,dex, which is about three times higher than that of the upper envelope value for the young stars in \citet{Romano2021}.  As discussed above, only  UCAC4\, 606-009417 is a wide binary system with an M1V component, therefore, it is not possible that these four stars enhance their Li by accreting the Li-rich material from its companion, such as AGB and RGB stars or novae. Considering all of them are fast rotators ($v\sin{i}>$15\,km\,s$^{-1}$), the most likely possibility is the accretion of circumstellar Li-rich matter.

There are three program stars with $v\sin{i}$ less than 10\,km\,s$^{-1}$, i.e., UCAC4\, 569-010383, 596-052739 and 677-057614,  and they have similar effective temperatures and  metallicities. All of them have very high Li abundnaces ($\mathrm{A(Li)_{NLTE}}$ $>$ 3.9\,dex), and there is no obvious evidence that the three stars are binaries based on the variation of their RVs and light curves from TESS. \citet{Koch2011} and \citet{Koch2012} suggested when a star engulfs a planet, its surface abundances of some elements, including Li, can elevate. More recently, \citet{Soares-Furtado2021} found that the most compelling strengths and survival times of engulfment-derived enrichment are the host stars near the main-sequence turn-off of mass between 1.4 to 1.6\,$M_{\sun}$.  However, the composition of the gaseous giant planet is similar to that of its host star, thus we can not expect this mechanism to enhance the Li abundance to their observed levels. Meanwhile, their Li enrichment by accreting the circumstellar Li-rich matter can not be ruled out.

UCAC4\, 629-030411 is of very fast rotation ($v\sin{i}=$ 45\,km\,s$^{-1}$). Its Li is more likely enriched by the accretion of circumstellar matter. However, this star has the highest temperature. It is suggested by \citet{Richer1993} and \citet{Richard2005} that Li from deeper regions can be radiatively accelerated outward to enrich the surface convection zone (diffusion) for stars in the temperature ranges of 6,900\,K and 7,100\,K. Considering the temperature of this star is only slightly lower than the above range, the possibility of diffusion enrichment can not be ruled out.

UCAC4\, 646-050572 is the most metal-poor star in our sample. The kinematic information of (U, V, W) = ($-$96, $-$120, $-$1) (km\,s$^{-1}$) indicates that it belongs to the thick-disk \citep{Bensby2018}. Considering its rotation velocity is around 11\,km s$^{-1}$ and no evidence of binary, the Li enrichment is most likely due to the accretion of circumstellar matter as well.

\section{Conclusions}\label{sec:conclusions}
We discover nine super Li-rich unevolved stars with NLTE Li abundance A(Li) higher than 3.8\,dex from the LAMOST-MRS, which is the largest sample of such type stars. It is found that the NLTE effects are as large as $\sim$0.5\,dex for these stars. We note that most of our program stars are fast rotator, therefore, the accretion of circumstellar matter may be the main contributor of the Li enhancement in these stars, however, other sources can not be ruled out. 

The LAMOST-MRS is ongoing, and will continue to provide a great opportunity to study such chemically peculiar stars.

\vspace{7mm} \noindent {\bf ACKNOWLEDGEMENTS}
The authors are grateful to Guang-Wei Li for helpful discussions about peculiar star. Our research is supported by National Key R\&D Program of China No.2019YFA0405502, the National Natural Science Foundation of China under grant Nos. 12090040, 12090044, 11833006, 12022304, 11973052, 11973042, U2031144 and U1931102.  H.-L.Y. acknowledges support from the Youth Innovation Promotion Association of the Chinese Academy of Sciences (id. 2019060) amd NAOC Nebula Talents Program. We acknowledge the science research grants from the China Manned Space Project with NO.CMS-CSST-2021-B05. Guoshoujing Telescope (the Large Sky Area Multi-Object Fiber Spectroscopic Telescope LAMOST) is a National Major Scientific Project built by the Chinese Academy of Sciences. Funding for the project has been provided by the National Development and Reform Commission. LAMOST is operated and managed by the National Astronomical Observatories, Chinese Academy of Sciences. This work has made use of data from the European Space Agency (ESA) mission Gaia (https://www. cosmos.esa.int/gaia), processed by the Gaia Data Processing and Analysis Consortium (DPAC, https://www. cosmos.esa.int/web/gaia/dpac/consortium). Funding for the DPAC has been provided by national institutions, in particular the institutions participating in the Gaia Multilateral Agreement. This paper includes data collected by the \emph{TESS} mission, which are publicly available from the Mikulski Archive for Space Telescopes (MAST). Funding for the \emph{TESS} mission is provided by NASA's Science Mission directorate.

\bibliography{ms}{}
\end{document}